\documentclass[journal]{IEEEtran}
\IEEEoverridecommandlockouts
\usepackage{cite}
\usepackage{breakurl}
\usepackage{amsmath,amssymb,amsfonts}
\usepackage{algorithmic}
\usepackage{graphicx}
\usepackage{textcomp}
\usepackage{xcolor}
\usepackage{subcaption}
\usepackage{empheq}
\def\BibTeX{{\rm B\kern-.05em{\sc i\kern-.025em b}\kern-.08em
    T\kern-.1667em\lower.7ex\hbox{E}\kern-.125emX}}
\usepackage[left=0.65in,top=0.76in,right=0.65in,bottom=0.76in]{geometry}

\makeatletter
\let\old@ps@headings\ps@headings
\let\old@ps@IEEEtitlepagestyle\ps@IEEEtitlepagestyle
\def\confheader#1{
	\def\ps@IEEEtitlepagestyle{
		\old@ps@IEEEtitlepagestyle
		\def\@oddhead{\normalfont\scriptsize\centering#1}%
	}%
}

\makeatother

\confheader{%
	This paper has been accepted for publication
	in IEEE International Conference on Computer Communications Workshops (INFOCOM Workshops), 29 April - 2 May 2019, Paris, France
}

\begin{document}

\title{An Energy-driven Network Function Virtualization for Multi-domain Software Defined Networks}

\author{
	\IEEEauthorblockN{Kuljeet Kaur\IEEEauthorrefmark{1}, Member, IEEE, Sahil Garg\IEEEauthorrefmark{1}, Member, IEEE, Georges Kaddoum\IEEEauthorrefmark{1}, Member, IEEE, \\Fran\c{c}ois Gagnon\IEEEauthorrefmark{1}, Senior Member, IEEE, Neeraj Kumar\IEEEauthorrefmark{2}, Senior Member, IEEE, and \\Syed Hassan Ahmed\IEEEauthorrefmark{3}, Senior Member, IEEE}\\
	\IEEEauthorblockA{\IEEEauthorrefmark{1}Electrical Engineering Department, \'Ecole de technologie sup\'erieure, Montr\'eal, QC H3C 1K3, Canada.\\
		\IEEEauthorrefmark{2}CSED, Thapar Institute of Engineering \& Technology, Patiala (Punjab), India.\\
		\IEEEauthorrefmark{3}Department of Computer Science, Georgia Southern University, Statesboro, GA 30460, USA.\\}
	(e-mail: kuljeet.kaur@ieee.org, sahil.garg@ieee.org, georges.kaddoum@etsmtl.ca, francois.gagnon@etsmtl.ca, neeraj.kumar@thapar.edu, and sh.ahmed@ieee.org)
}
	\maketitle
	\begin{abstract}
		
	Network Functions Virtualization (NFV) in Software Defined Networks (SDN) emerged as a new technology for creating virtual instances for smooth execution of multiple applications. Their amalgamation provides flexible and programmable platforms to utilize the network resources for providing Quality of Service (QoS) to various applications. In SDN-enabled NFV setups, the underlying network services can be viewed as a series of virtual network functions (VNFs) and their optimal deployment on physical/virtual nodes is considered a challenging task to perform. However, SDNs have evolved from single-domain to multi-domain setups in the recent era. Thus, the complexity of the underlying VNF deployment problem in multi-domain setups has increased manifold.	
	Moreover, the energy utilization aspect is relatively unexplored with respect to an optimal mapping of VNFs across multiple SDN domains. 	
	Hence, in this work, the VNF deployment problem in multi-domain SDN setup has been addressed with a primary emphasis on reducing the overall energy consumption for deploying the maximum number of VNFs with guaranteed QoS. The problem in hand is initially formulated as a \textit{``Multi-objective Optimization Problem"} based on Integer Linear Programming (ILP) to obtain an optimal solution. However, the formulated ILP becomes complex to solve with an increasing number of decision variables and constraints with an increase in the size of the network. Thus, we leverage the benefits of the popular evolutionary optimization algorithms to solve the problem under consideration. In order to deduce the most appropriate evolutionary optimization algorithm to solve the considered problem, it is subjected to different variants of evolutionary algorithms on the widely used MOEA framework (an open source java framework based on multi-objective evolutionary algorithms). The experimental results demonstrate that the proposed scheme achieves better results in comparison to the $\epsilon$-Non-dominated Sorting Genetic Algorithm (NSGA)-II ($\epsilon$-NSGA-II) with the respect to the overall energy consumption and optimal deployment of VNFs in multi-domain SDN scenarios.
	
	\end{abstract}
	
	

\section{Introduction} \label{sec:Introduction}
With the development of new agile technologies such as 5G (and beyond), the number of connected devices are increasing at a rapid pace. However, the lack of direct application program interface (API) interactions with network services impose limitations on the handling of the rapidly changing traffic patterns. Also, the rise of cloud, Internet of Things (IoT) and AI-driven applications have made this task more challenging. To cope with the ever-increasing demands of networks in a cost-efficient manner, the conventional networks are demanding more flexible and dynamic solutions \cite{singh2018bloom, 8422407}. Towards this end, network softwarization is gaining huge popularity which supports rapid provisioning and deployment of network services to meet QoS requirements \cite{kaur2018edge}. By reducing overhead, automating infrastructure and increasing maintainability, it aims to build agile and flexible networks to provision convergent networking solutions.\\
\indent In this direction, Software Defined Networks (SDN) and Network Functions Virtualization (NFV) have been introduced \cite{7892961}. SDN is an exciting and evolving paradigm that turns the network into a flexible and programmable platform able to optimally utilize the resources. It separates the control plane of a network from the data plane to make it centrally manageable \cite{8613868}. NFV, on the other hand, represents the network service as a series of virtual network functions (VNFs). In NFV architectures, individual VNFs can be chained together to perform the desired sequence, known as service function chain (SFC). Using SFC, a large number of VNFs can be connected together in an NFV environment. To support a fully virtualized network, VNFs provide network scalability and agility, while enabling better provisioning of network resources \cite{7243304}. \\
\indent {To meet the dynamic requirements of the network resources, several solutions have been proposed in the literature for SFCs deployment. For example, Chen \textit{et al.} \cite{8570806} proposed a fully decentralized online approach for service chaining based on stochastic dual-gradient method. Similarly, Sun \textit{et al.} \cite{8423711} studied the problem of SFC orchestration using a mesh aggregation approach where a feedback mechanism was deployed to improve its success rate. In another work, Sun \textit{et al.} \cite{8565965} proposed a heuristic algorithm using a restricted Boltzmann machine and cross-entropy approach for low-latency applications in NFV networks. Likewise, Gupta \textit{et al.} \cite{GUPTA20181} proposed a VNF service chain placement model to facilitate the dynamic service demands of network traffic. Despite the rapid progression, the dynamic composition of services is still facing imminent challenges which need immediate attention. In this direction, the integration of SDN and NFV has been proposed as a viable solution; where NFV provides the basic functionality to the network, while SDN controls and orchestrates them for specific uses. Thus, the convergence of both technologies facilitates the flexible deployment of networks and service delivery over them. \\
	\indent In the similar context, Medhat \textit{et al.} \cite{medhat2017service} explored the limitations of current SFC approaches in next-generation networks and showed that approaches integrating SDN and NFV technologies have higher SFC scalability and flexibility compared to others. To ensure the benefits of the service-chaining policy, Ding \textit{et al.} \cite{7113222} proposed OpenSCaaS platform by integrating SDN and NFV. Furthermore, Zhang \textit{et al.} \cite{8450542} presented a framework to construct efficient SFCs by combining SDN and NFV.  Bruschi \textit{et al.} \cite{8063920} proposed a softwarized SDN/NFV approach to support the cloud services in a more scalable and sustainable way. Likewise, Pie \textit{et al.} \cite{8392781} developed a novel routing algorithm to solve the SFC problem in SDN and NFV-enabled network. In order to minimize the resource consumption flow costs with SFC requests, binary integer programming was used. Although several attempts have been made to tackle the service deployment of networks, they all rely on single domain SDN networks. However, the rapidly growing network traffic and constant enlargement of network scales pose challenges to single domain SDN architectures in terms of latency, overhead, fault tolerance and load balancing, thereby affecting the overall QoS experience for end-users. \\
	\indent To overcome these challenges, large-scale SDN networks have been proposed. However, the deployment of network services on a large-scale SDN network involves the distribution of VNFs on multiple SDN domains. Thus, the service provisioning across these networks requires the collaboration among different SDN controllers which results in additional network overhead. However, the deployment of network services across multi-domain SDNs is still in its early phases \cite{zhang2018network}. Further, the deployment of network services in multi-domain SDN networks requires an efficient allocation of resources for different service requests at a reduced cost. Thus, it is of primary interest to implement energy efficient solutions which dynamically adapt the usage of resources according to the demands.

	\subsection{Contributions}
	This paper proposes an energy-efficient approach to deploy NFVs in multi-domain SDN networks. To the best of our knowledge, the evaluation of energy characteristics in multi-domain SDN setups is a novel research initiative and has not been explored. The key contributions of the proposed wok are summarized as follows:
	
	\begin{enumerate}
		\item We analyze the problem of VNF deployment in multi-domain SDN environments (associated with energy consumption and resource allocation challenges), and mathematically formulate a multi-objective optimization problem for the effective deployment of NFV in SDN. The formulated problem aims at achieving an optimal mapping of VNFs across a multi-domain SDN setup while minimizing the overall energy consumption level. 
		\item We then evaluate the proposed multi-objective problem on different evolutionary algorithms to have an optimal solution in polynomial time using the widely used MOEA framework. To attain this objective, different variants of Non-dominated Sorting Genetic Algorithm (NSGA) are used namely NSGA-II, NSGA-III and $\epsilon-$NSGA-II. 
		\item We then extensively assess the performance of the proposed scheme using the most appropriate evolutionary algorithm under different scenarios in terms of number of VNFs, number of SDN domains, number of constraints and decision variables. 
	\end{enumerate}
	
	\subsection{Organization}
	The rest of this paper is organized as follows. Section~\ref{sec:SystemModel} presents the system model for the proposed scheme. In Section~\ref{sec:ProblemFormulation}, the problem of VNF deployment in multi-domain SDN is formulated. Finally, in Section~\ref{sec:ResultsAndDiscussions}, the formulated problem is subjected to different variants of NSGA in different networks scenarios. Finally, this work is concluded in Section~\ref{sec:ConclusionAndFutureScope} with future directions.
\section{System Model} \label{sec:SystemModel}

In the following section, we detail the system model for the next-generation communication systems empowered by SDN and NFV. The designed solution helps in the optimized mapping of the VNFs to the available SDN nodes (scattered across multiple domains) keeping the energy consumption and resource utilization profile in check.

\begin{figure}[h]
	\centering
	\includegraphics[scale=.58]{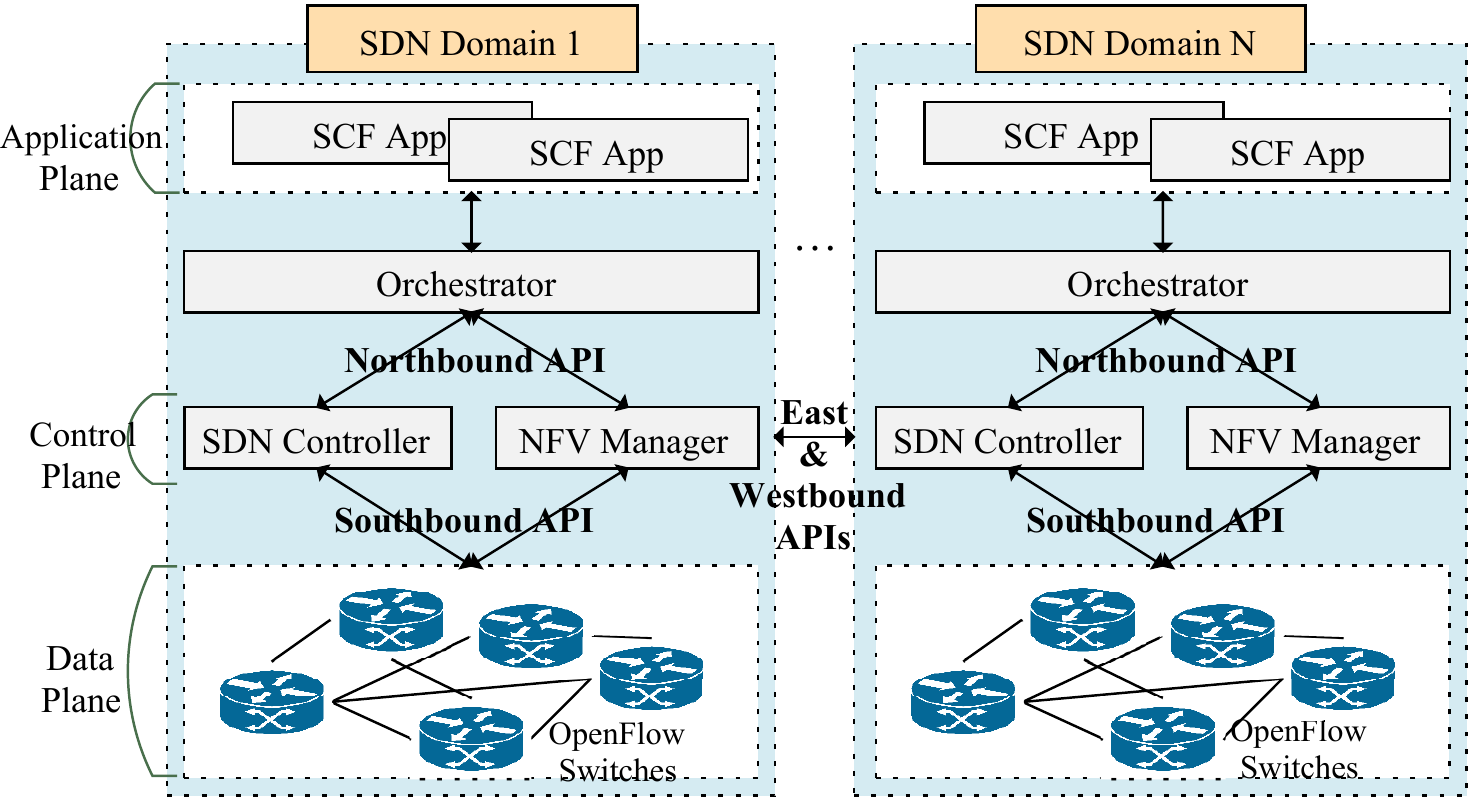}
	\caption{Architectural model based on SDN and NFV.}
	\label{fig:systemmodel}
\end{figure}
\noindent Fig.~\ref{fig:systemmodel} depicts the system model of the proposed communication system. It represents the coupling of SDN with NFV for seamless networking operations. 
SDN is a software-based programming paradigm for modern networks that rely on software reconfigurations for changing the network policies. As shown in the figure, the SDN platform is comprised of a data plane and a control plane. Here, the data plane consists of the Open-Flow (OF)-enabled routers and switches with dedicated flow table entries. These OF-flow tables are primarily controlled by the SDN's control plane via the southbound API, \textit{i.e.}, OpenFlow protocols. In other words, the policy changes are assumed by the SDN's controller; which are reproduced at the data plane \cite{chaudhary2017network}.


On the other hand, NFV is an upcoming networking paradigm that helps to segregate the network functions from the underlying proprietary hardware. Further, it helps to execute the network functions on software with flexible implementation which are referred to as VNFs. By coupling NFV with the powerful paradigm of SDN, the network services can be established with ease by redirecting the traffic through the appropriate VNFs in a sequential manner. The Internet Engineering Task Force (IETF) defines the abstract notion of VNFs using  SFC--a means to implement VNFs. These SFCs are managed by the SDN's centralized controller and deployed on the underlying hardware. With the SDN-NFV coupling, the deployment of SFCs becomes flexible without the need to change the hardware.
 \begin{figure}[!ht]
	\centering
	\includegraphics[scale=.57]{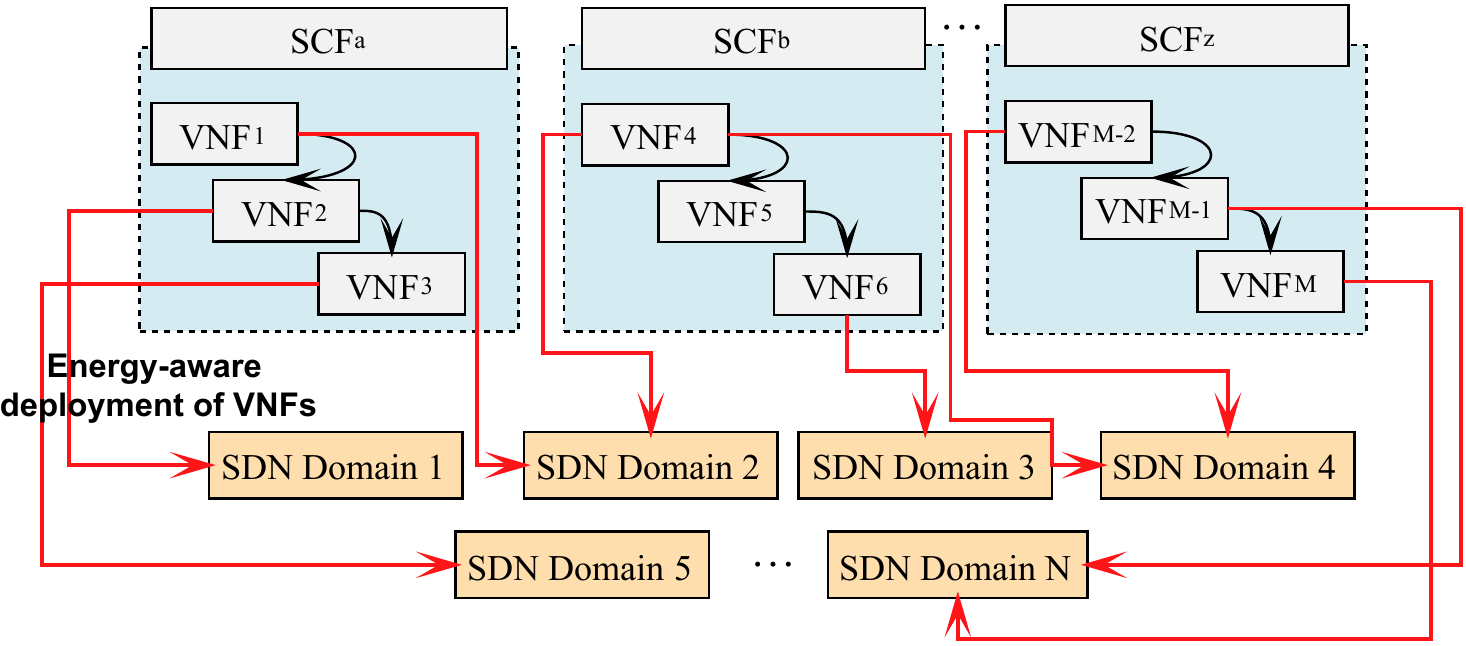}
	\caption{Deployment of VNFs across SDN domains.}
	\label{fig:systemmodelworking}
\end{figure}
\newline However, the constant enlargement of networks has resulted in wide-scale SDNs with several interconnected domains referred to as multi-domain SDNs \cite{zhang2018network}. All the domains are individually administrated by their respective controllers and are linked with each other to provide the required network services. The provision of these network services involves the successful deployment of VNFs (belonging to different SCFs) across different SDN domain nodes (physical or virtual). Nevertheless, the collaboration between different SDN domain controllers is essential for the efficient distribution of resources amongst the VNFs. The same can be achieved using east and west-bound APIs. In summary, the focus of the proposed work is to manage the underlying resources effectively while attaining minimal energy consumption. In the proposed work, the same is achieved by mapping VNFs to the available array of SDN domains while assigning the required resources (CPU cores, memory and storage) to VNFs for their execution. The details of the overall deployment process are illustrated using Fig.~\ref{fig:systemmodelworking}.

\section{Problem Formulation} \label{sec:ProblemFormulation}
In this section, the problem of VNF deployment across a multi-domain SDN set-up is illustrated. The key objective functions of the considered problem followed by their corresponding constraints are presented.

\vspace{2mm}
\noindent \textit{A. List of Objective Functions}
\vspace{1mm}

As discussed earlier, the proposed scheme supports two major objective functions ($\mathfrak{F}(\mathbb{X}_{ij})$) as illustrated under. Here, $\mathbb{X}_{ij}$ denotes the decision variable expressed as follows:
\begin{IEEEeqnarray}{rc}
	\mathbb{X}_{ij} =&  \begin{cases}
		1: \text{If $i^{th}$ VNF is deployed on the $j^{th}$ SDN domain}\\
		0: \text{Otherwise}
	\end{cases}
	\label{eq:DecisionVariable}
\end{IEEEeqnarray}
In the above equation, the nature of $\mathbb{X}_{ij}$ is binary, \textit{i.e.}, $\mathbb{X}_{ij} \in \{0,1\}$. Here, $\mathbb{X}_{ij}=1$ indicates that the $i^{th}$ VNF is deployed successfully on the $j^{th}$ SDN domain, while $\mathbb{X}_{ij}=0$ suggests that $i^{th}$ VNF is not deployed on the $j^{th}$ SDN domain.

\vspace{2mm}
\indent \textit{1) Maximum deployment of VNFs:} One of the foremost objectives of the proposed scheme is to ensure maximum deployment of VNFs on the available array of SDN nodes. This is attained by using a maximization operation over the summation of $\mathbb{X}_{ij}$ as follows:
\begin{equation}
\mathfrak{F}_1(\mathbb{X}_{ij}) = \max \sum_{i=1}^{\mathrm{M}} \sum_{j=1}^{\mathrm{N}}  \mathbb{X}_{ij} 
\label{eq:ObjectiveFunction1}
\end{equation}
In the above equation, $\mathrm{M}$ denotes the total number of VNFs to be deployed.

\vspace{2mm}
\indent \textit{2) Minimum Energy Consumption:}
This objective function deals with the minimization of the overall energy consumption involved in executing VNFs on the available SDN domains. It is denoted  as follows:
\begin{equation}
\mathfrak{F}_2(\mathbb{X}_{ij}) = \min \sum_{i=1}^{\mathrm{M}} \sum_{j=1}^{\mathrm{N}} \mathbb{X}_{ij} \times \mathbb{E}_{ij}
\label{eq:ObjectiveFunction2}
\end{equation}
where, $\mathrm{N}$ denotes the total number of SDN domains and variable $\mathbb{E}_{ij}$ refers to the energy consumption (in kWh) associated with the execution of the $i^{th}$ VNF over the $j^{th}$ SDN domain.

\vspace{3mm}
\noindent \textit{B. List of Constraints}
\vspace{1mm}

This section presents different set of constraints that should be respected while achieving energy-aware deployment of VNFs in the multi-domain SDN setup.  

\vspace{2mm}
\indent \textit{1) VNF deployment constraint:}
The following constraint ensures that only a single VNF is allocated at most a single SDN domain as expressed using the below mentioned equation. The said constraint restricts the multiple deployment of an individual VNF in multi-domain SDN environments and helps to minimize the overall energy consumption.
\begin{IEEEeqnarray}{rcl}
	\mathrm{C}_1: & \mathbb{X}_{i1} + \mathbb{X}_{i2} + \dots + \mathbb{X}_{i\mathrm{N}} = 1; & \forall i \nonumber, \\
	\implies & \sum_{j=1}^{\mathrm{N}} \mathbb{X}_{ij}= 1; &\forall i. \label{eq:Constraint1}
\end{IEEEeqnarray}

\vspace{2mm}
\indent \textit{2) Resource restriction with respect to CPU cores:}
The following constraint ensures that VNF deployment on different physical nodes always meets their required CPU core needs. The said restriction helps to allocate the requested number of cores to the mapped VNFs using the following equation. In other words, the constraints ensure that no more VNFs can be mapped to a physical node without the required number of CPU cores requested by the corresponding VNF. 
\begin{IEEEeqnarray}{rcl}
	\mathrm{C}_{2}: & \sum_{i=1}^{} \mathbb{X}_{1j} \times \mathbb{C}_1 \dots +  \mathbb{X}_{\mathrm{M}j} \times \mathbb{C}_\mathrm{M} \le \mathfrak{C}_j; & \forall j, \nonumber \\
	\implies & \sum_{i=1}^{\mathrm{M}} \mathbb{X}_{ij} \times \mathbb{C}_i \le \mathfrak{C}_j; &\forall j. \label{eq:Constraint4}
\end{IEEEeqnarray}
In the above equation, $\mathbb{C}_i$ refers to the number of the CPU cores required by the $i^{th}$ VNF and $\mathfrak{C}_j$ denotes the upper limit on the $j^{th}$ node's CPU core capacity. 

\vspace{2mm}
\indent \textit{3) Resource restriction with respect to Memory:}
In line, with the above mentioned constraint, the following restriction imposes the upper limit of the main memory that could be allocated to the VNFs on a particular SDN domain node.
\begin{IEEEeqnarray}{rcl}
	\mathrm{C}_{3}: & \sum_{i=1}^{} \mathbb{X}_{1j} \times \mathbb{M}_1 \dots +  \mathbb{X}_{\mathrm{M}j} \times \mathbb{M}_\mathrm{M} \le \mathfrak{M}_j; & \forall j, \nonumber \\
	\implies & \sum_{i=1}^{\mathrm{M}} \mathbb{X}_{ij} \times \mathbb{M}_i \le \mathfrak{M}_j; &\forall j. \label{eq:Constraint2}
\end{IEEEeqnarray}
here, variables $\mathbb{M}_i$ and $\mathfrak{M}_j$ represent the main memory required by the $i^{th}$ VNF and the upper limit on the main memory capacity of the $j^{th}$ node, respectively.

\vspace{2mm}
\indent \textit{4) Resource restriction with respect to Storage:}
The following equation represents the resource restriction corresponding to disk storage allocation.
\begin{IEEEeqnarray}{rcl}
	\mathrm{C}_{4}: & \sum_{i=1}^{} \mathbb{X}_{1j} \times \mathbb{S}_1 \dots +  \mathbb{X}_{\mathrm{M}j} \times \mathbb{S}_\mathrm{M} \le \mathfrak{S}_j; & \forall j, \nonumber \\
	\implies & \sum_{i=1}^{\mathrm{M}} \mathbb{X}_{ij} \times \mathbb{S}_i \le \mathfrak{S}_j; &\forall j. \label{eq:Constraint3}
\end{IEEEeqnarray}
The variable $\mathbb{S}_i$ defined in the above equation denotes the disk storage requested by the $i^{th}$ VNF; while the variable $\mathfrak{S}_j;$ denotes the upper limit on disk storage for the $j^{th}$ node.

\vspace{3mm}
\noindent \textit{C. Overall Problem at hand}
\vspace{1mm}

\noindent The overall problem can be summarized as  follows:
\begin{IEEEeqnarray}{rCl}
	\mathfrak{F}(\mathbb{X}_{ij})  = &   \big[-\mathfrak{F}_1(\mathbb{X}_{ij}),& \mathfrak{F}_2(\mathbb{X}_{ij})\big] \label{eq:MOOP}\\
	\text{such that} & & \nonumber\\
	\text{Constraints in} &
	\begin{cases}
		\text{~Eq.~}\eqref{eq:Constraint1} \\ 
		\text{~Eq.~}\eqref{eq:Constraint4} \\ 
		\text{~Eq.~}\eqref{eq:Constraint2} \\ 
		\text{~Eq.~}\eqref{eq:Constraint3} \\ 
		
	\end{cases}
	& \text{~holds}  \nonumber
\end{IEEEeqnarray}

\noindent As detailed above, the considered problem comprises of the two objective functions $\mathfrak{F}_1(\mathbb{X}_{ij})$ and $\mathfrak{F}_2(\mathbb{X}_{ij})$. The minus sign before $\mathfrak{F}_1(\mathbb{X}_{ij})$ denotes maximization operation, \textit{i.e.}, the aim is to maximize the deployment of VNFs. On the other hand, $\mathfrak{F}_2(\mathbb{X}_{ij})$ depicts the minimization operation over the energy consumption objective function. The associated constraints span from Eq.~\eqref{eq:Constraint1} till Eq.~\eqref{eq:Constraint3}. It is worth noticing here that the nature of the problem mentioned above is linear with linear constraints; which suggests that the problem under consideration is ILP with binary decision variable $\mathbb{X}_{ij}$. However, solving an ILP with a huge number of parameters (say thousands of $\mathbb{X}_{ij}$) is quite a gigantic task. This is because with increasing Internet traffic, multi-domains SDNs are expected to grow exponentially, which in turn would lead to a substantial increase in the number of VNFs ($\mathrm{M}$) and SDN-domains ($\mathrm{N}$). Consider the following example,
\begin{empheq}[box=\fbox]{align*}
\mathrm{M} & =  100;  \nonumber\\
\mathrm{N} & =  20;  \nonumber \\
\therefore  \text{~No. of Decision Variables}& ~(|\mathbb{X}_{ij}|)=  \mathrm{M} \times \mathrm{N} \nonumber\\
& = 100 * 20= 2000 \nonumber\\
\text{~No. of Constraints}& ~(|\mathrm{C}_{k}|)=  \mathrm{M} + 3 \times \mathrm{N} \nonumber\\
& =  100 + 3* 20= 160 \nonumber
\end{empheq}

\vspace{1mm}

\noindent Thus, with the constant increase in parameters, an optimal solution for the above problem would be time-consuming and impossible in polynomial time. 
Henceforth, the proposed scheme leverages the advantages of a popular multi-objective evolutionary algorithm called $\epsilon$-NSGA-II. It is characterized by its fast convergence, low overall complexity, explicit diversity preservation capability, epsilon-dominance, adaptive population sizing, and automatic termination \cite{reed2003simplifying, deb2002fast}.

 The experimental evaluation against the existing state-of-the-art multi-objective evolutionary algorithms demonstrate that $\epsilon$-NSGA-II is best suited for our problem, as justified by the preliminary results demonstrated in the upcoming section.

\section{Results and Discussions} \label{sec:ResultsAndDiscussions}
This section illustrates the experimental evaluation of the proposed scheme compared to different state-of-the-art schemes on different evaluation parameters.

To evaluate the proposed VNF deployment, the widely accepted MOEA Framework has been employed \cite{moeaframework}.  It is an open source Java library which is heavily used by the research community for designing and evaluating different multi-objective evolutionary algorithms such as $\epsilon$-NSGA-II, NSGA-II, NSGA-III, generalized differential evolution (GDE3), strength pareto evolutionary algorithm 2 (SPEA2), indicator-based evolutionary algorithm (IBEA), speed-constrained multi-objective particle swarm optimization (SMPSO) covariance matrix adaptation evolution strategy (CMA-ES), multi-objective evolutionary algorithm with decomposition (MOEA/D), and $\epsilon$-MOEA. Additionally, it also supports rapid design and development of customized problems and algorithms. For the proposed scheme, the formulated multi-objective optimization problem (defined Eq.~\eqref{sec:ProblemFormulation}) has been scripted in java using MOEA Framework and has been evaluated with existing state-of-the-art algorithms such as NSGA-II, NSGA-III, and $\epsilon$-NSGA-II.  

\begin{figure}
	\begin{subfigure}[b]{0.45\textwidth}
		\centering
		\includegraphics[scale=.9]{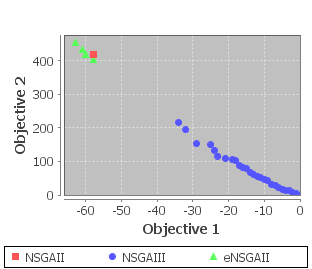}
		\caption{Pareto front evaluation.}
		\label{fig:n2}
	\end{subfigure}
	
	\begin{subfigure}[b]{0.45\textwidth}
		\centering
		\includegraphics[scale=.9]{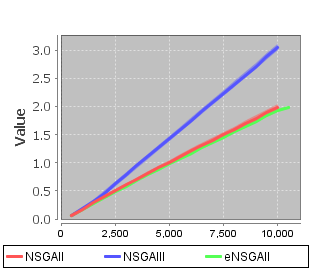}
		\caption{Elapsed time evaluation.}
		\label{fig:n1}
	\end{subfigure}
	\caption{An illustrative comparison of NSGA-II, NSGA-III and $\epsilon$-NSGA-II for solving the problem under consideration.}
	\label{fig:MOEAComparison}
\end{figure}

The experiment evaluation for the proposed multi-objective scheme was carried out across two phases (referred to as Case Studies). In the first phase of the assessment, the proposed scheme was subjected to different multi-objective evolutionary algorithms, and their relative performance was compared considering Pareto front and elapsed time. Based on the comparison, it was concluded that $\epsilon$-NSGA-II was the most appropriate algorithm to address the proposed NFV deployment across multi-domain SDN setups. Accordingly, in the next phases, the efficacy of the proposed scheme was evaluated using $\epsilon$-NSGA-II on a dataset spanning different scenarios. The detailed description of these phases is illustrated below.

\begin{figure}
	\centering
	\begin{subfigure}[b]{0.4\textwidth}
		\centering
		\includegraphics[scale=.45]{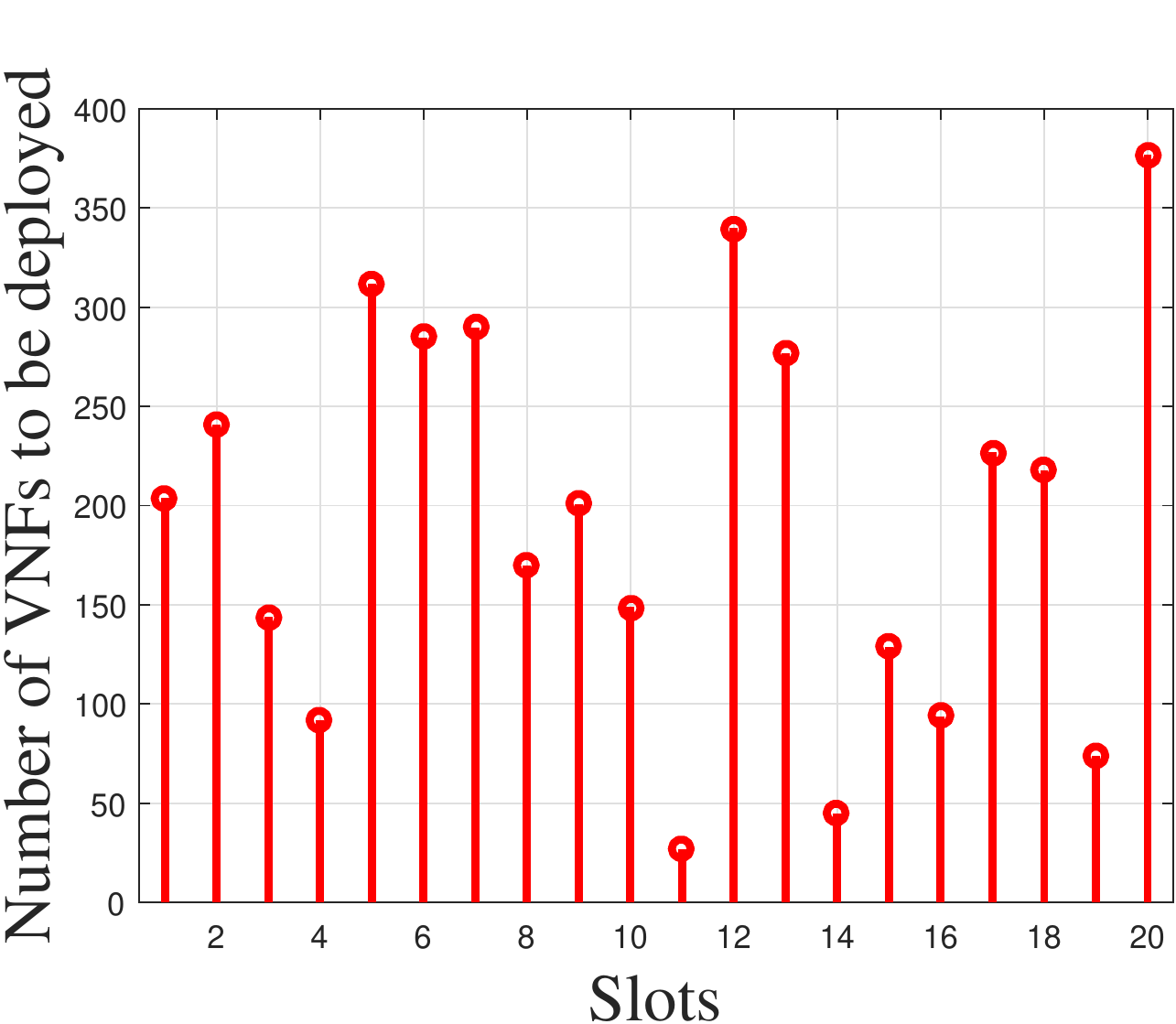}
		\caption{Number of VNFs to be deployed.}
		\label{fig:vnfstobedeployed}
		\medskip\smallskip
	\end{subfigure}

	\begin{subfigure}[b]{0.4\textwidth}
		\centering
		\includegraphics[scale=.45]{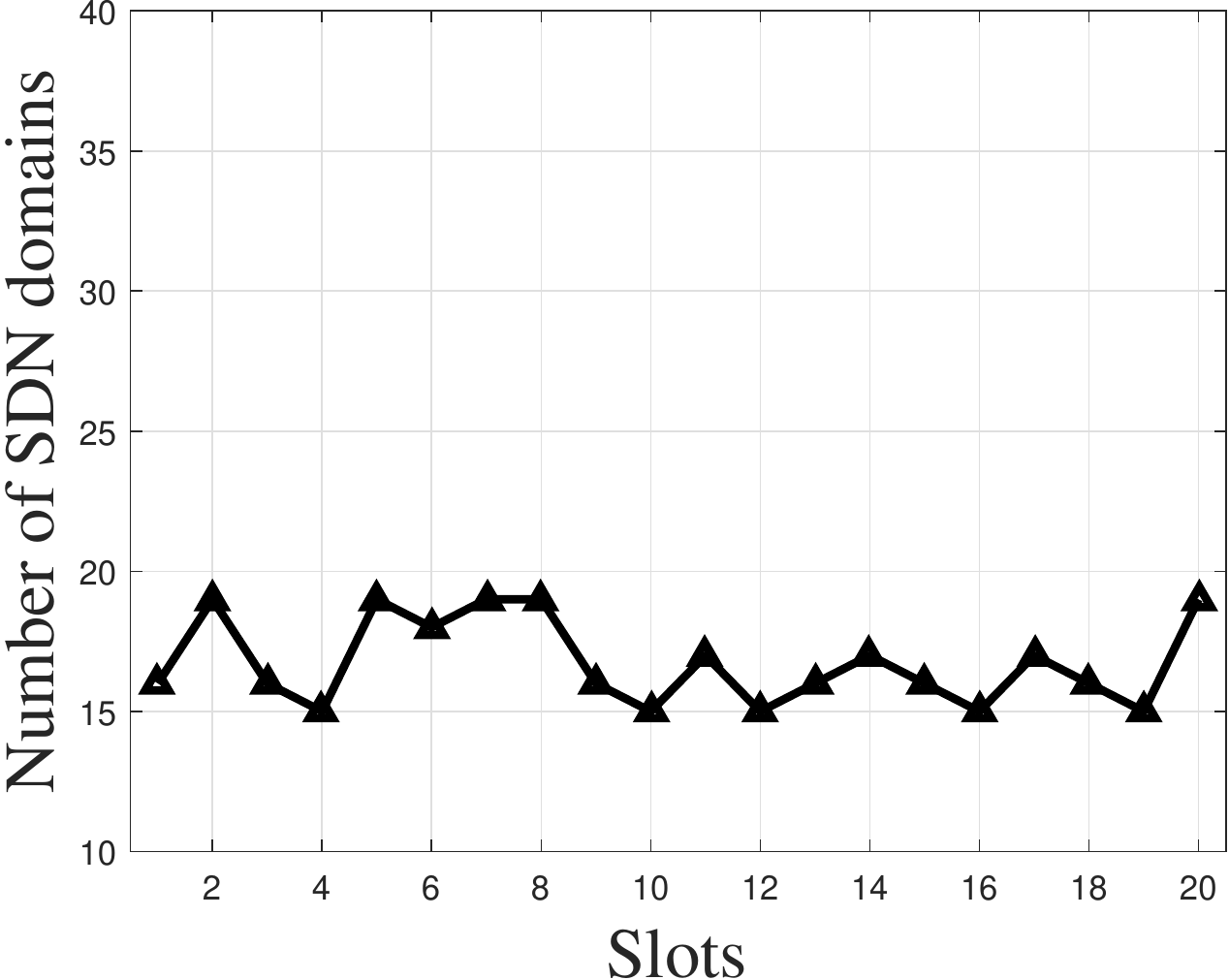}
		\caption{Number of SDN domain nodes.}
		\label{fig:sdndomains}
		\medskip\smallskip
	\end{subfigure}
	
	\begin{subfigure}[b]{0.4\textwidth}
		\centering
		\includegraphics[scale=.45]{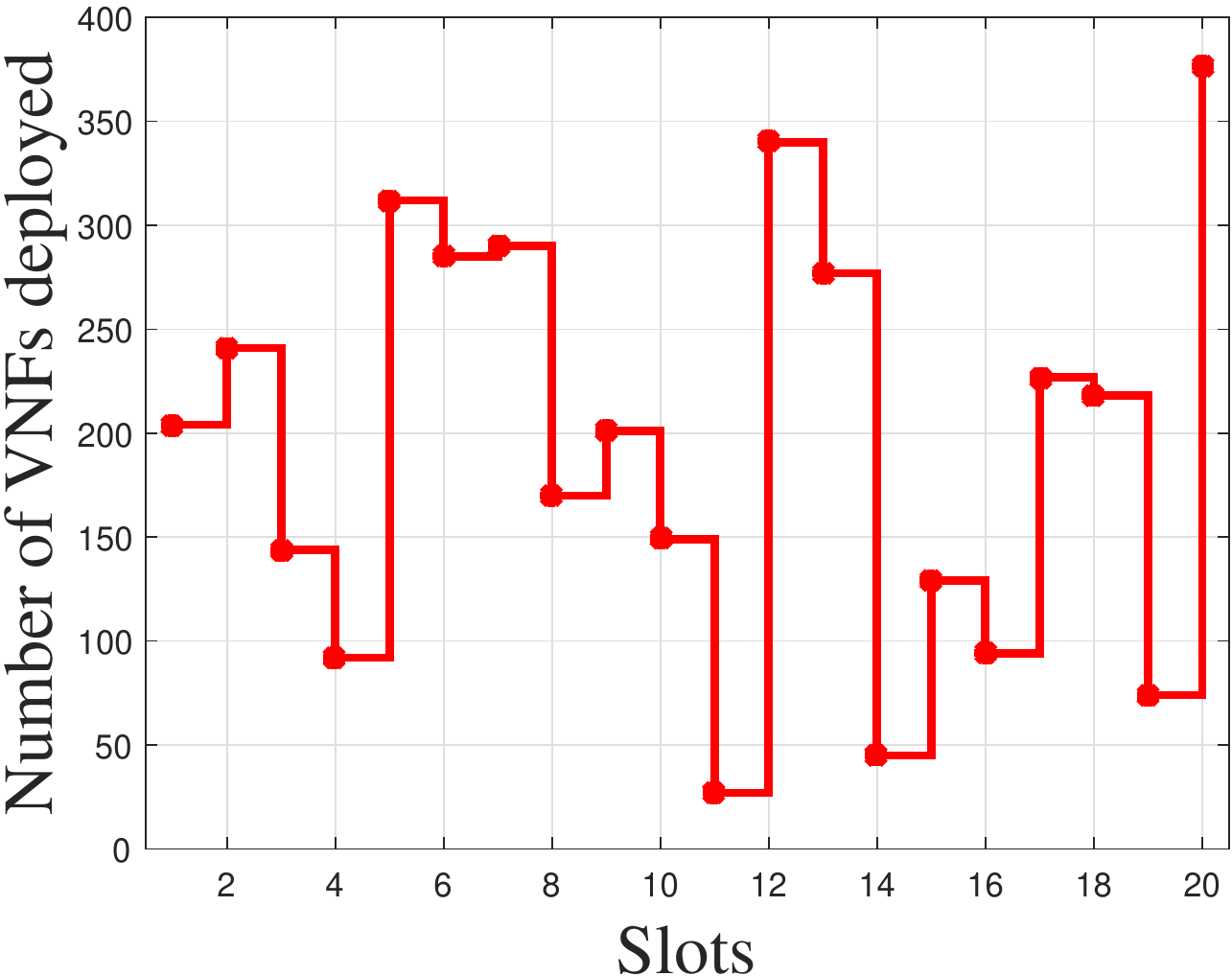}
		\caption{Execution of $\mathfrak{F}_1$ for successful deployment of VNFs.}
		\label{fig:numberofdeployements}
		\medskip\smallskip
	\end{subfigure}
	
	\begin{subfigure}[b]{0.4\textwidth}
		\centering
		\includegraphics[scale=.45]{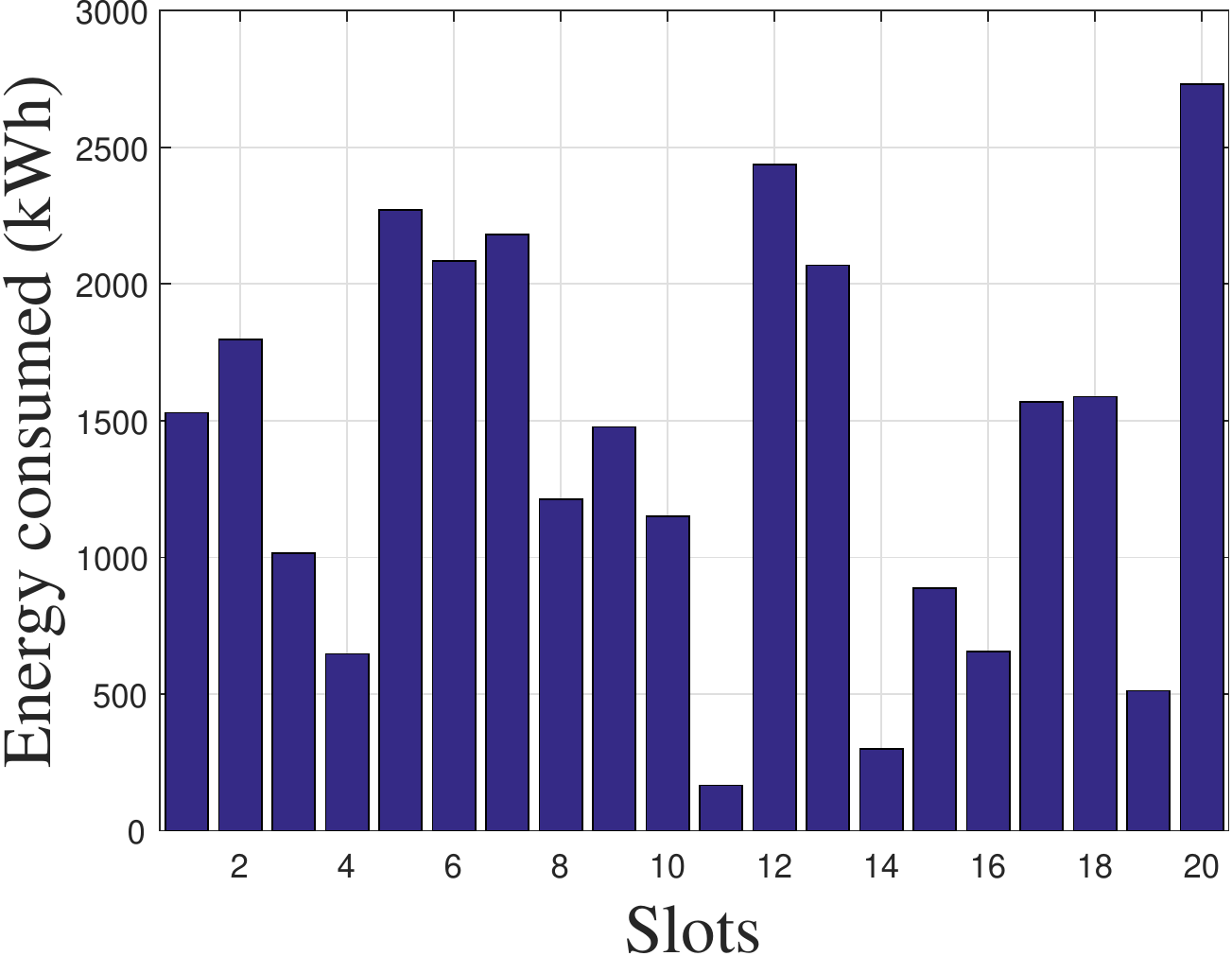}
		\caption{Execution of $\mathfrak{F}_2$ for minimizing energy consumption.}
		\label{fig:energycosnumed}
		\medskip\smallskip
	\end{subfigure}
	
	\caption{A illustration of executing the proposed multi-objective optimization problem for VNF deployment in multi-domain SDN using $\epsilon$-NSGA-II.}\label{fig:ResultsOverall}
\end{figure}

\noindent \textit{A. Case Study I}
\vspace{1mm}

The problem of VNF deployment in multi-domain environments has been evaluated using different versions of NSGA namely NSGA-II, NSGA-III, and $\epsilon$-NSGA-II. The related results are summarized in Fig.~\ref{fig:MOEAComparison}. Each of these algorithms has been evaluated with 10,000 function evaluations with 20 seeds in the MOEA framework's Diagnostic Tool. The related results regarding pareto front and elapsed time are summarized in Fig.~\ref{fig:n2} and Fig.~\ref{fig:n1}, respectively. The pareto front
evaluations show that $\epsilon$-NSGA-II achieves the most optimal performance followed by NSGA-II and NSGA-III; in terms of trade-off between the two objective function $\mathfrak{F}_1(\mathbb{X}_{ij})$ and $\mathfrak{F}_2(\mathbb{X}_{ij})$. Although the performance of $\epsilon$-NSGA-II and NSGA-II is almost identical; however the former results in much less energy consumption deployments as compared to the other. Additionally, the computational time of the three algorithms is summarized in Fig.~\ref{fig:n1}. Similar to the previous case, $\epsilon$-NSGA-II  achieves the optimal performance with a reduced value of elapsed time. This suggests that $\epsilon$-NSGA-II achieves the optimal performance with quick computational processing.

Thus, for both parameters, $\epsilon$-NSGA-II stands out which makes it the ideal choice to solve the considered VNF deployment challenge in a multi-domain SDN setup.

\vspace{3mm}
\noindent \textit{B. Case Study II}
\vspace{1mm}

During this case study, the performance of the proposed multi-objective approach for VNF deployment has been evaluated using $\epsilon$-NSGA-II under different scenarios in terms of number of VNFs deployed, number of SDN domain nodes, number of decision variables and number of constraints. The variability concerning the first two parameters can be understood from the data plotted in Figs.~\ref{fig:vnfstobedeployed} and Fig~\ref{fig:sdndomains}. For the other two parameters, the following equations can be used for the computational purpose.
\begin{IEEEeqnarray}{rcl}
	\text{~No. of Decision Variables}& ~(|\mathbb{X}_{ij}|)=  &\mathrm{M} \times \mathrm{N} \nonumber\\
	\text{~No. of Constraints}& ~(|\mathrm{C}_{k}|)= & \mathrm{M} + 3 \times \mathrm{N} \nonumber
\end{IEEEeqnarray}
In order to achieve successful deployment of VNFs in a multi-domain setup, Eq.~\eqref{eq:MOOP} was evaluated using $\epsilon$-NSGA-II. The related results in terms of successful VNF deployment and energy consumptions statistics are highlighted in Fig.~\ref{fig:numberofdeployements} and Fig.~\ref{fig:energycosnumed}, respectively. The obtained results evidently support the fact that the proposed scheme achieves an optimal trade-off between the two competing objective functions $\mathfrak{F}_1(\mathbb{X}_{ij})$ and $\mathfrak{F}_2(\mathbb{X}_{ij})$. Overall, in comparison with the existing NSGA variants, the proposed approach using $\epsilon$-NSGA-II gives better performance regarding energy consumption and successful VNF deployment.

\section{Conclusion} \label{sec:ConclusionAndFutureScope}

Gradual evolutions in the networking infrastructures have resulted in the convergence of SDN and NFV to provide seamless deployment of network services. Additionally, this convergence has led to the amplification of SDNs from single domain to multiple domains. Under such multi-domain SDNs, the realization of NFV would become a challenging task particularly due to the complex interactions between SDN controllers of different domains.  Particularly, the energy consumption management will also evolve as a challenging task. Thus, in this paper, the problem of NFV deployment in multi-domain SDNs has been addressed. The main focus of the proposed work was to minimize the energy consumption caused by VNF deployment across multiple domain SDN while achieving their optimal deployment with the required number of resources. In summary, the said problem has been modeled as an ILP and solved using different variants of NSGA-a popular evolutionary  optimization algorithm. However, experimental evaluations suggests that $\epsilon$-NSGA-II gives the optimal trade-off between the considered objective functions and is best suited for the problem under consideration. Additionally, the experimental results also suggest that the  proposed VNF deployment solution also reduces the energy consumption to a large extent.

In the near future, we will extend this work to locality-based VNF deployment; where VNFs of a particular SFC would be deployed in a specific SDN domain.

%
%
%
%
%

\bibliographystyle{IEEEtran}
\bibliography{RefIntro.bib,Ref.bib}

\end{document}